\documentclass{article}
\usepackage{spconf,amsmath,graphicx}
\usepackage{amssymb}
\usepackage{multirow}
\usepackage{booktabs}
\usepackage{ctable}

\title{MONAURAL SPEECH ENHANCEMENT WITH Complex convolutional block attention module and joint time frequency losses}
\name{Shengkui Zhao, Trung Hieu Nguyen, Bin Ma}
\address{Speech Lab, Alibaba Group}
\begin{document}
%
\maketitle
\begin{abstract}
Deep complex U-Net structure and convolutional recurrent network (CRN) structure achieve state-of-the-art performance for monaural speech enhancement. Both deep complex U-Net and CRN are encoder and decoder structures with skip connections, which heavily rely on the representation power of the complex-valued convolutional layers. In this paper, we propose a complex convolutional block attention module (CCBAM) to boost the representation power of the complex-valued convolutional layers by constructing more informative features. The CCBAM is a lightweight and general module which can be easily integrated into any complex-valued convolutional layers. We integrate CCBAM with the deep complex U-Net and CRN to enhance their performance for speech enhancement. We further propose a mixed loss function to jointly optimize the complex models in both time-frequency (TF) domain and time domain. By integrating CCBAM and the mixed loss, we form a new end-to-end (E2E) complex speech enhancement framework. Ablation experiments and objective evaluations show the superior performance of the proposed approaches.
\end{abstract}
\section{Introduction}
\label{sec:intro}

Speech enhancement is a highly desired task in speech communication applications where the perceptual quality and intelligibility can be severely decreased by the additive background noise. The goal of speech enhancement is to extract the signal of interest from the corrupted noisy speech for better perceptual quality and intelligibility.

For decades, monaural speech enhancement has been a challenging problem. Recently, deep learning methods have made significant performance improvement over conventional signal processing based methods. Given clean speech and background noise, the speech enhancement task can be formulated as a supervised learning problem with the simulated noisy speech as input and the clean speech as target. Although researchers have tried to build deep learning models directly on time-domain speech signals \cite{Pascual2017, Rethage2018, Luo2019}, it is more common to work on time-frequency (TF) representations via short-time-Fourier-transform (STFT). Comparing to the spectral mapping approach \cite{Ronneberger2015}, the TF mask-based approach has been more popular \cite{Srinivasan2006, Narayanan2013, Wang2014}. Early studies usually focus on enhancing the magnitude spectrum while reusing the noisy phase spectrum. Recent studies show the phase and magnitude components are relative important in terms of speech perceptual quality and intelligibility, leading researchers to consider both magnitude and phase estimation. The phase-sensitive mask (PSM) \cite{Erdogan2015} was one of the first attempts to incorporate phase information into mask estimation. However, it still estimates a real component.  Subsequently, the ideal complex ratio mask (CRM) \cite{Williamson2015} was proposed to estimate both the real and the complex components and it theoretically achieves the best oracle speech enhancement performance.  The pioneer work \cite{Williamson2015} attempts to estimate CRM via real-valued network operations, which are not coincided well with the complex-valued masks. More recently, deep complex networks \cite{Trabelsi2018} were developed and the complex-valued models \cite{Choi2019, Isik2020, Hu2020} achieve state-of-the-art (SOTA) performance for speech enhancement. We found that the SOTA models \cite{Choi2019, Isik2020, Hu2020} are built on either the U-Net structure \cite{Ronneberger2015} or the convolutional recurrent network (CRN) structure \cite{Tan2018}, which heavily rely on the representation power of the convolutional layers. The attention mechanisms \cite{Hu2018, Woo2018} proposed for image processing that fuse both spatial and channel-wise information to boost the representation power of the convolutional layers have not been exploited in the SOTA complex models for speech enhancement.

In this paper, we present a lightweight and general complex-valued channel-spatial attention mechanism, built upon previous studies \cite{Woo2018} to boost the representation power of the complex-valued convolutional layers, therefore enhance capabilities of the complex-valued SOTA models, e.g., U-Net and CRN, for speech enhancement. Our attention mechanism is formed as an individual module called complex convolutional block attention module (CCBAM) which can be easily integrated into any complex-valued convolutional layers with negligible overheads and is end-to-end trainable along with base network. The CCBAM can be interpreted as a means of biasing the allocation of available computational resources towards the most informative features for optimizing representation power. Our experiments suggest adding CCBAM to both the decoder layers and the skip connections provide better results and achieve superior performance. Besides the proposed attention mechanism, we also add mean-squared error (MSE) loss of the real and the complex components of CRM to the scale-invariant source-to-noise ratio (SI-SNR) loss \cite{Choi2019, Hu2020} to jointly optimize the CRM estimation in the TF domain and the time domain. Our experiments will demonstrate the effectiveness of the proposed joint loss functions for speech enhancement.

The related works are described as follows. The attention blocks named self-attention were applied for speech enhancement \cite{Isik2020, Koizumi2020}. The self-attention focused on the response of positions in a sequence while CCBAM focuses on cross-channel and spatial information of feature maps. Another related work is the spatial-channel gating scheme proposed for the U-Net structure \cite{Khanh2019}, which addressed medical image segmentation problem. \cite{Lan2020} also applied a concurrent space-channel-wise attention to the redundant convolutional encoder-decoder (RCED) for speech enhancement. However, their work was based on magnitude spectrum mapping. The above mentioned works were based on real-valued networks and we focus on complex-valued networks.

\section{The Complex Convolutional Block Attention Module}
\label{sec:format}

Our proposed CCBAM is a refined complex-valued attention mechanism applied in STFT-domain based on the work described in \cite{Woo2018}. It is composed of a complex channel-attention module and a complex spatial-attention module as shown in Fig. 1 and Fig. 2. Both modules consist of a squeeze operation and an excitation operation. The squeeze operation involves pooling process to aggregate information. The excitation process applies the aggregated information to generate recalibration weights.
\begin{figure}[t]
  \centering
  \includegraphics[width=\linewidth]{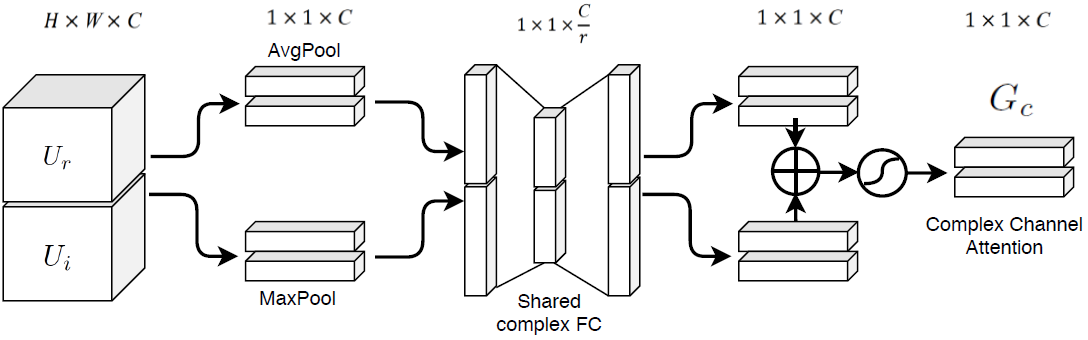}
  \caption{Diagram of complex channel attention module.}
  \label{fig1}
\end{figure}

\begin{figure}[t]
  \centering
  \includegraphics[width=6cm]{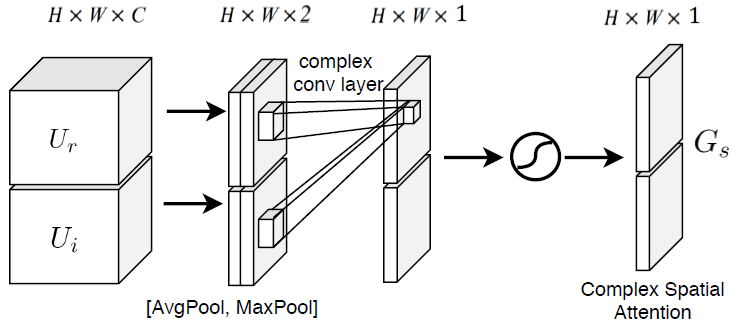}
  \caption{Diagram of complex spatial attention module.}
  \label{fig1}
\end{figure}
\subsection{Complex channel-attention module}
Let the output complex feature map be $U=U_r+jU_i$ where $U\in\mathbb{C}^{H\times W \times C}$ from the complex-valued convolutional filter $W=W_r+jW_i$ with complex input matrix $V=V_r+jV_i$. The complex convolution operation can be formulated as:
\begin{align}
U_r = V_r \ast W_r - V_i \ast W_i \nonumber \\
U_i = V_r \ast W_i + V_i \ast W_r
\end{align}
where $\ast$ stands for real-valued convolutional filtering. Applying global average pooling to $U_r$ and $U_i$ to aggregate spatial information of the feature map, respectively, we obtain $U_r^\mathrm{avg}=\mathrm{AvgPool}(U_r)$, $U_i^\mathrm{avg}=\mathrm{AvgPool}(U_i)$. Similarly, applying max pooling we obtain $U_r^\mathrm{max}=\mathrm{MaxPool}(U_r)$, $U_i^\mathrm{max}=\mathrm{MaxPool}(U_i)$. Then, we apply two complex-valued FC layers to $U^\mathrm{avg}=U_r^\mathrm{avg}+jU_i^\mathrm{avg}$ and $U^\mathrm{max}=U_r^\mathrm{max}+jU_i^\mathrm{max}$, respectively. Let the complex weights of the two FC layers be $W^\mathrm{FC}_1=W^\mathrm{FC}_{1r}+jW^\mathrm{FC}_{1i}$ and $W^\mathrm{FC}_2=W^\mathrm{FC}_{2r}+jW^\mathrm{FC}_{2i}$, where $W^\mathrm{FC}_1 \in \mathbb{C}^{\frac{C}{r}\times C}$ and $W^\mathrm{FC}_2 \in \mathbb{C}^{C \times \frac{C}{r}}$, with $r$ standing for a reduction ratio. The complex-valued channel-attention gate $G_c \in \mathbb{R}^{1\times 1\times C}$ is then computed as:
\begin{equation}
G_c = \sigma(W^\mathrm{FC}_2\delta(W^\mathrm{FC}_1U^\mathrm{avg})) + \sigma(W^\mathrm{FC}_2\delta(W^\mathrm{FC}_1U^\mathrm{max}))
\end{equation}
where $\delta$ refers to complex-valued ReLU function and $\sigma$ refers to complex-valued sigmoid function \cite{Trabelsi2018}, and both ReLU and sigmoid apply on real and imaginary values, respectively. In Eq. (2), the weights $W^\mathrm{FC}_1$ and $W^\mathrm{FC}_2$ are shared between $U^\mathrm{avg}$ and $U^\mathrm{max}$. The complex FC operations can be performed in a similar way as the complex convolution in Eq. (1).

\subsection{Complex spatial-attention module}
Applying average-pooling and max-pooling operations to $U_r$ and $U_i$ to aggregate channel information of the feature map $U$, respectively, and then applying a complex-valued two-dimensional (2D) convolutional filter $F=F_r+jF_i$ to the concatenated pooling ouput, we obtain the complex spatial-attention gate $G_s \in \mathbb{R}^{H\times W \times 1}$ as:
\begin{equation}
G_s = \sigma(F[\mathrm{AvgPool}(U);\mathrm{MaxPool}(U)])
\end{equation}
where $\sigma$ denotes the complex-valued sigmoid function and $[\mathrm{AvgPool}(U);\mathrm{MaxPool}(U)]$ represents the concatenation operation along the channel axis. The pooling operations are performed on $U_r$ and $U_i$ separately. The complex convolution operation works similarly as defined in Eq. (1). The convolution filter size is a hyperparameter and we set to $7 \times 7$ for better performance in our experiments.

\begin{figure}[t]
  \centering
  \includegraphics[width=\linewidth]{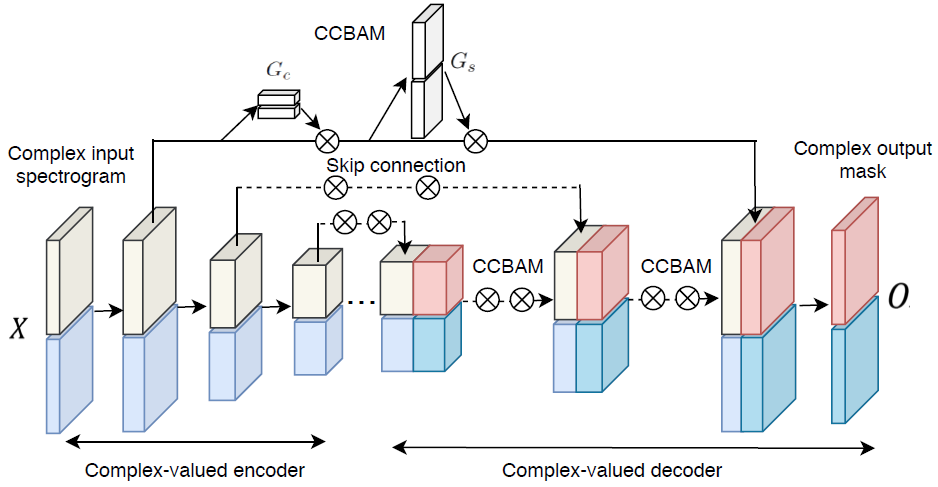}
  \caption{Illustration of CCBAM integrated with convolutional encoder-decoder structure with skip connections. The symbol $\otimes$ stands for element multiplication.}
  \label{fig1}
\end{figure}

\begin{figure}[t]
  \centering
  \includegraphics[width=\linewidth]{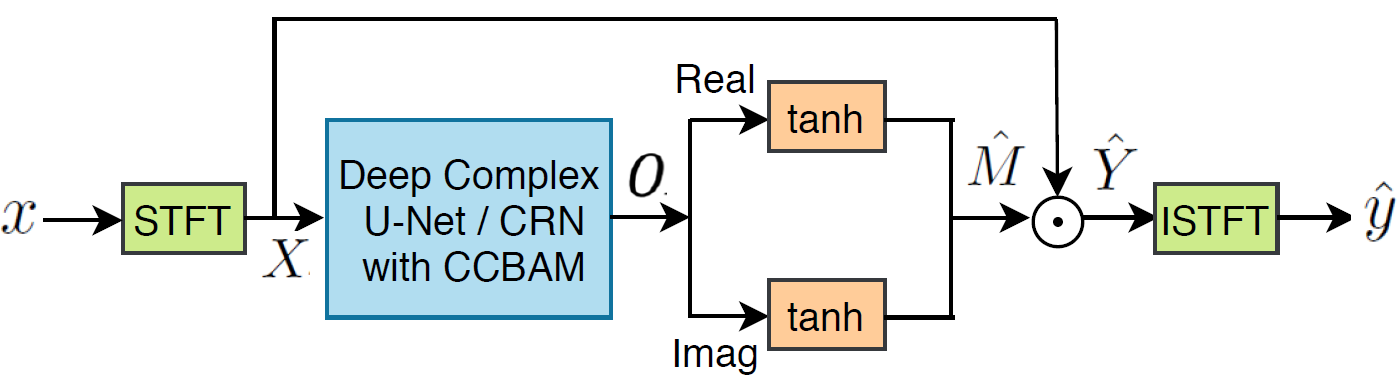}
  \caption{Illustration of the proposed end-to-end speech enhancement method.}
  \label{fig1}
\end{figure}

\section{Integration of CCBAM with Deep Complex U-Net and CRN Architectures}
\label{sec:pagestyle}
\subsection{Integration of CCBAM}
The deep complex U-Net (DCUnet) \cite{Choi2019} is a complex-valued network design of the real-valued U-Net structure proposed in \cite{Ronneberger2015} for image segmentation. The DCUnet is composed of a complex-valued convolutional encoder and a decoder with skip-connections as shown in Fig. 3. The encoder gradually extracts high-lever features with a down-sampling process and the decoder reconstructs the target map with an up-sampling process. The encoder and the decoder use complex 2D convolutional and deconvolutional layers, respectively. Each complex convolutional/deconvolutional operation is followed by complex batch normalization and leaky ReLU activation. The skip-connections connect the convolutional layers from the encoder to the corresponding deconvolutional layers of the decoder. The deep complex CRN (DCCRN) \cite{Hu2020} is a complex-valued network design of the real-valued CRN structure proposed in \cite{Tan2018}. Comparing to the DCUnet, the DCCRN is also composed of a complex-valued convolutional encoder and a decoder with skip-connections. The essential difference is that the DCCRN added LSTM layers between the encoder and the decoder aiming to model the temporal dependencies. Since the DCUnet and DCCRN share equivalent encoder and decoder structure, we integrate CCBAM into DCUnet and DCCRN similarly. Fig. 3 shows the overall integrated architecture of the convolutional encoder-decoder network with our proposed CCBAM. We insert the two complex channel and spatial attention modules of CCBAM into the decoder paths and the skip connection paths in a sequential manner. The CCBAM performs as self-gating mechanism to recalibrate the TF feature maps adaptively in consideration of the global information. It can help the skip connections to by-pass meaningful information that are beneficial for CRM estimation. Meanwhile, the CCBAM is also expected to help the decoder to concentrate on 'what' and 'where' of feature maps when performing target reconstruction.
\subsection{The end-to-end speech enhancement method}
The proposed end-to-end speech enhancement method is illustrated in Fig. 4. Assuming the noisy speech signal is denoted as $x(n)=y(n)+z(n) \in \mathbb{R}$ where $y(n)$ is the clean speech signal and $z(n)$ the background noise, the speech enhancement task is to estimate $y(n)$ from $x(n)$. Let $X=X_r+jX_i \in \mathbb{C}$, $Y=Y_r+jY_i \in \mathbb{C}$ denote the STFT representations of $x(n)$ and $y(n)$, respectively. The CRM method \cite{Williamson2015} is to estimate the complex-valued mask $M=Mr+jM_i \in \mathbb{C}$ such that $Y = M \odot X$, where $\odot$ indicates complex multiplication. The ground-truth CRM can be computed as:
\begin{equation}
M = \frac{X_r\cdot Y_r+X_i\cdot Y_i}{X^2_r+X^2_i}+j\frac{X_r\cdot Y_i-X_i\cdot Y_r}{X^2_r+X^2_i}.
\end{equation}
Estimating the unbounded components $M_r$ and $M_i$ is very challenging. In our speech enhancement model, we propose to add {\it tanh} function to bound the estimated real and imaginary components, $\hat{M}_r$ and $\hat{M}_i$, of $\hat{M} \in \mathbb{C}$. The estimated clean speech spectrogram $\hat{Y}$ is then obtained as
\begin{equation}
\hat{Y} = (X_r\cdot\hat{M}_r-X_i\cdot\hat{M}_i)+j(X_r\cdot\hat{M}_i+X_i\cdot\hat{M}_r).
\end{equation}
The estimated time domain clean speech $\hat{y}(n)$ is obtained by applying ISTFT on $\hat{Y}$.

In the SOTA frameworks \cite{Choi2019, Hu2020}, the neural models are optimized based only on the time domain loss function SI-SNR or weighted SI-SNR. The SI-SNR loss may not be fully correlated with the CRM estimation. We propose an improved mixed loss function by integrating SI-SNR loss with mean squared error (MSE) losses of the real and the imaginary CRM estimates. Specifically, we optimize the neural model by the following loss function
\begin{equation}
\mathcal{L}(y,\hat{y}) = \lambda_\mathrm{SI-SNR}\mathcal{L}_\mathrm{SI-SNR}(y,\hat{y})+\lambda_\mathrm{Mask}\mathcal{L}_\mathrm{Mask}(M,\hat{M}),
\end{equation}
where the time domain SI-SNR loss $\mathcal{L}_\mathrm{SI-SNR}(y,\hat{y})$ is defined as \cite{Hu2020}:
\begin{equation}
\mathcal{L}_\mathrm{SI-SNR}(y,\hat{y}) = \mathrm{10log10}\bigg(\frac{||y_\mathrm{target}||^2_2}{||e_\mathrm{noise}||^2_2}\bigg)
\end{equation}
where $y_\mathrm{target}$ is given as: $y_\mathrm{target}=(<\hat{y},y>\cdot y)/||y||^2_2$ and the residual noise $e_\mathrm{noise}$ is  $e_\mathrm{noise}=\hat{y}-y_\mathrm{target}$. Here, $<\cdot,\cdot>$ denotes dot product and $||\cdot||_2$ is L2 norm.
The mask loss function is defined as
\begin{equation}
\mathcal{L}_\mathrm{Mask}(M,\hat{M}) = \sum_{t,f}[(\hat{M}_r-M_r)^2+(\hat{M}_i-M_i)^2]
\end{equation}
where $t,f$ are omitted for simplicity.
The STFT and ISTFT operations are implemented
as 1-D convolutional and deconvolutional layers to facilitate the end-to-end training.
\section{Experiments}
\label{sec:typestyle}
\subsection{Datasets}
We evaluated and compared the proposed methods on two datasets. The dataset-1 is a relative small set where we selected about 50 hours clean speech from WSJ0 \cite{Garofolo1993} and about 50 hours noise from DEMAND \cite{Thiemann2013} and  RNNoise \footnote{https://media.xiph.org/rnnoise}. The clean speech includes 131 speakers and the noise includes various types of computer fans, office, crowd, airplane, car, train, construction, etc. We used 40 hours of clean speech and 40 hours of noise for training and validation, and the rest for test. The noisy speech mixture was obtained by randomly mixing speech utterances with noise segments. The noise segments were split or concatenated to meet the length of speech utterances. A wide range of SNRs between 0 dB and 20 dB were randomly generated for both training and evaluation. The total training data was about 100 hours and the total evaluation data was about 30 minutes.

The dataset-2 is a relative big set provided by the DNS challenge \cite{Reddy2020}. It contains over 500 hours clean speech and 180 hours noise selected from various public resources. For the noisy speech mixture, we used the SNR range between -5 dB to 20 dB. The clean speech and the noise were randomly mixed and we generated a total of 2000 hours training data. Note that we targeted only clean speech and background noise in this work and leave the reverberant case for future work. We evaluated on the official synthetic test set of the DNS challenge. All waveforms were resampled at 16k Hz.
\subsection{Training setup}
For the DCUnet model, we used the same model configuration as the DCUnet-20 shown in Fig. 7 in \cite{Choi2019}. The DCUnet-20 has 20 convolutional layers and the total number parameters are 3.5M. We replaced the original SI-SNR loss of the DCUnet-20 with our proposed mixed loss and adopted the model into our end-to-end method as shown in Fig. 4. The resulted model was denoted as DCUnet-M. We further applied our proposed CCBAM into DCUnet-M and denoted the model as DCUnet-MC. We also followed the same pre-processing in \cite{Choi2019} to apply STFT with a 64ms sized Hann window and 16ms hop length. For the DCCRN model, we used the real-time CRN model configuration as described in \cite{Tan2018}. The original CRN model was proposed for magnitude spectrum mapping. We extended it to complex-valued network as described in \cite{Hu2020} with 20M trainable parameters. Similar to DCUNet, we integrated DCCRN with our proposed mixed loss and CCBAM. The resulted models were denoted as DCCRN-M and DCCRN-MC, respectively. For DCCRN, the window length and hop size were 20 ms and 10 ms. For all model training, we used Pytorch platform and Adam optimizer. We set learning rate to 0.001 and decay it by 0.5 when the validation score increases. For the loss parameters, we set $\lambda_\mathrm{SI-SNR} = 0.5$ and $\lambda_\mathrm{Mask}=0.5$.

\begin{table}
\caption{Average performance evaluation results for different models on dataset-1. }
\begin{tabular}{lcccc}
\specialrule{.1em}{.05em}{.05em}
\multirow{2}{*}{Model} & \multicolumn{4}{c}{Evaluation  Metrics}         \\
\cline{2-5}
                       & PESQ   & STOI   & SI-SNR   & FwSegSNR  \\ \hline
Noisy                  & 1.97   & 87.83  & 6.28     & 10.64     \\ \hline
DCCRN                  & 3.17   & 95.80  & 17.71    & 21.35     \\
DCCRN-M                & 3.28   & 96.30  & 18.03    & 23.06     \\
DCCRN-MC               & 3.34   & 96.76  & 18.42    & \textbf{24.18}     \\ \hline
DCUnet                 & 3.25   & 96.26  & 18.31    & 20.16     \\
DCUnet-M               & 3.38   & 96.85  & 18.59    & 22.10     \\
DCUnet-MC              & \textbf{3.44}   & \textbf{97.10}  & \textbf{19.33}    & 23.81     \\ \hline
\specialrule{.1em}{.05em}{.05em}
\end{tabular}
\end{table}

\begin{table}
\caption{Average performance evaluation results for different models on dataset-2. }
\begin{tabular}{lcccc}
\specialrule{.1em}{.05em}{.05em}
\multirow{2}{*}{Model} & \multicolumn{4}{c}{Evaluation Metrics}         \\
\cline{2-5}
                       & PESQ   & STOI   & SI-SNR   & FwSegSNR  \\ \hline
Noisy                  & 2.45   & 91.52  & 9.17     & 13.55     \\
DNS Baseline           & 2.68   & 77.62  & -26.63   & 7.39      \\ \hline
DCCRN                  & 3.04   & 93.95  & 15.70    & 19.77     \\
DCCRN-M                & 3.15   & 94.87  & 16.03    & 19.92     \\
DCCRN-MC               & 3.21   & 95.10  & 16.50    & \textbf{20.65}     \\ \hline
DCUnet                 & 3.20   & 94.80  & 16.48    & 19.53     \\
DCUnet-M               & 3.24   & 95.88  & 16.92    & 19.63     \\
DCUnet-MC              & \textbf{3.30}   & \textbf{96.11}  & \textbf{17.46}    & 20.31     \\ \hline
\specialrule{.1em}{.05em}{.05em}
\end{tabular}
\end{table}
\subsection{Results}
For the objective evaluation, we used 4 evaluation metrics: PESQ, STOI, SI-SNR, and FwSegSNR \cite{Hu2008}. Table 1 and 2 show objective metric evaluation results on dataset-1 and dataset-2, respectively. From the results, we can see that the DCUnet-M and DCCRN-M outperform the original baseline DCUnet and DCCRN. This shows the proposed mixed loss is beneficial to the CRM estimation. Moreover, we can see that DCUnet-MC and DCCRN-MC outperforms their corresponding models without attention mechanism. This shows the benefits of our proposed CCBAM for model optimization.
\section{Conclusions}
In this paper, we presented a complex convolutional block attention module (CCBAM) and a mixed loss function to improve the deep complex U-Net and CRN models for monaural speech enhancement. The integrated end-to-end complex speech enhancement framework with the deep complex U-Net and CRN models achieve better performance compared to their original models. We conducted experimental studies and demonstrated the superior performance of the proposed methods on two different datasets.
\bibliographystyle{IEEEbib}
\bibliography{refs}

\end{document}